\documentclass[useAMS,usenatbib]{mn2e}
\usepackage{graphicx}

\title[Complex emission line region of Mrk 817]
  {Complex emission line region of Mrk 817}
\author[D. Ili\'c et al.]
  {D.~Ili\'c,$^1$\thanks{{\it{Send offprint requests to:}} dilic@matf.bg.ac.yu}
  L.\v C.~Popovi\'c,$^2$ E.~Bon,$^2$
 E.G.~Mediavilla$^3$, V.H.~Chavushyan$^{4, 5}$
   \\
  $^1$Department of Astronomy, Faculty of Mathematics,
University of Belgrade, Studentski trg 16, 11000 Belgrade, Serbia\\
  $^2$Astronomical Observatory, Volgina 7, 11000 Belgrade, Serbia\\
  $^3$Instituto de  Astrof\'isica de Canarias, C/ V\'ia L\`actea, s/n
E38200 La Laguna (Tenerife), Spain\\
  $^4$Instituto Nacional de Astrofisica, Optica y Electronica,
  Apartado Postal 51, CP 72000 Puebla, Pue. M\'exico\\
  $^5$Instituto de Astronom\'{\i}a, UNAM,
Apartado Postal 70-264, 04510 M\'exico D.F., M\'exico
      }
\date{Released 2006 Xxxxx XX}

\pagerange{\pageref{firstpage}--\pageref{lastpage}} \pubyear{2006}

\def\LaTeX{L\kern-.36em\raise.3ex\hbox{a}\kern-.15em
    T\kern-.1667em\lower.7ex\hbox{E}\kern-.125emX}

\def\kms{\ifmmode {\rm km\ s}^{-1} \else km s$^{-1}$\fi}
\def\deg{\ifmmode ^{\rm o} \else $^{\rm o}$\fi}
\def\cm3{\ifmmode {\rm cm}^{-3} \else cm$^{-3}$\fi}

\begin{document}

\label{firstpage}

\maketitle

\begin{abstract}
In this work we study the physical and kinematical properties of the
emission line region of Seyfert 1.5 galaxy Mrk 817 using three sets
of observations, among which are high-resolution spectra obtained
with the Isaac Newton Telescope on Canary Islands. We find that in
Mrk 817 the Narrow (NEL) and Broad Emission Lines (BEL) are very
complex, indicating that structure of both the Narrow (NLR) and
Broad Line Region (BLR) is {\bf complex} and consists of at least
two sub-regions with different kinematical properties. We find that
the BEL can be fitted with the two-component model, where the core
of the line is coming from a spherical region with isotropic
velocity distribution, and wings might be affected by a low inclined
accretion disc (or disc-like emitting region). Also, we discuss the
physical properties of the BLR. Moreover, we find that an outflow is
present in the NLR, that may be driven by an approaching jet.
\end{abstract}


\section{Introduction}

The emission line region of Active Galactic Nuclei (AGN) is complex
and is usually divided to the Narrow (NLR) and Broad Line Region
(BLR), where the physics of the NLR is better understood then that
of the BLR (see e.g. Sulentic et al. 2000). The most accepted
scenario of the structure of AGN is the one in which AGN are powered
by the accretion of matter from the host galaxy on to super-massive
black hole. One of the way to study the inner emitting region of an
AGN, one that is closest to the black hole, is by analyzing its
broad emission lines. So far, in a small fraction of AGN
double-peaked emission lines were detected (around 5 \%, see
Eracleous \& Halpern~2003). Modeling these lines gave proof to the
presence of an accretion disc in the AGN (Eracleous \&
Halpern~1994,~2003). Beside the disc, emission lines also imply
presence of more kinematically different emission regions that
contribute to formation of lines: complex broad and narrow line
regions (Popovi\'c et al.~2003).

Even there are numerous papers devoted to the studies of the
kinematical and physical properties of the NLR and BLR (Krolik
~1999, Kembhavi \& Narlikar~1999, Sulentic et al.~2000), it is not
clear yet what is the connection between these two kinematically
different regions. One of the method to study the connection is to
map a whole emission region of an AGN. For that one should have a
high resolution spectra of the object that covers wide wavelength
band. It is also needed that the AGN emits both narrow and broad
lines. Accordingly, the spectroscopical investigations of Seyfert 1
galaxies that have strong narrow lines (as e.g. Seyfert 1.5) are
important. Following these reasons, for our analysis we selected Mrk
817, that is a Seyfert 1.5 galaxy, with the redshift of 0.03145,
{\bf (Strauss \& Huchra~1988) determined from the emission 
lines}\footnote{\bf A slightly different value of 0.031158, determined 
from both absorption and emission lines, can be found in 
Falco et al.~(1999), but since the redshift is not crucial for 
this study we adopt the value of 0.03145 for the systemic redshift 
of the galaxy.},
and which both broad and narrow emission lines are complex
(Popovi\'c \& Mediavilla~1997; Peterson et al.~1998; Popovi\'c et
al.~2004). Also, for this galaxy the mass ($M_{\rm BH}\approx 4.9
\times 10^7 M_{\odot}$) and the BLR size ($R_{\rm BLR}\approx 15$
light days\footnote{The radius of 15 light days corresponds to $3400
\ R_{\rm g}$, where $R_{\rm g}=GM/c^2$ is the gravitational radius
for the black hole mass of $4.9 \times 10^7 M_{\odot}$ (G is the
gravitational constant, M is the mass of the black hole and c is the
speed of light).}) have been estimated by reverberation mapping
studies (Peterson et al.~1998,~2004; Kaspi et al.~2000). We observed
the galaxy several times collecting the high resolution spectra in
the H$\alpha$ and H$\beta$ wavelength band, as well as the low
resolution spectra in the wide wavelength region.

The aim of this work is to explore the properties of the whole
emission region of the active galaxy Mrk 817. Using the
high-resolution spectra (such as one obtained with the Isaac Newton
Telescope) in analyzing the broad spectral line shapes and applying
the two-component model of the BLR proposed by Popovi\'c et al.
~(2004), we investigate the kinematical parameters of the BLR in Mrk
817. First we will present the Gaussian analysis of the H$\alpha$
and H$\beta$ lines and after that we will apply the two-component
model for fitting the broad emission lines. Also, an estimate of the
electron temperature in the BLR will be given. Finally we make a
scheme of the emission line region of Mrk 817 and discuss the
complex structure of that region.

\section{Observations and data reduction}

In studying the BLR of Mrk 817, three different sets of
spectral observations were used:

i) Observations with the 2.5-m Isaac Newton Telescope (INT) at La Palma
island in Spain. The observations were performed in the period
21--25 of January 2002. The Intermediate Dispersion Spectrograph
(IDS) and the 235 camera in combination with the R1200Y grating were
used. Two exposures of 550 and 500 s, included three H$\alpha$ and
three H$\beta$ spectrum. The seeing was $1''.1$ and the slit width
$1''$. The spectral resolution was around 1.0 \AA .

ii) Observations with the 4.2-m William Herschel Telescope (WHT), at
La Palma islands. The observations were performed on 12/13 March
2001. The long-slit spectrograph (ISIS) was used, in combination
with CCD cameras TEK4 (grating R158R). The H$\alpha$ was observed
with the exposure of 120 s.  The slit width of the spectrograph
was $0.8''$. The spectral resolution was about
2.9 \AA .

iii) Intermediate resolution optical spectroscopy was carried out
with Mexican 2.1-m telescope in Observatorio Astronomico Nacional de
San Pedro Martir (OAN SPM). Observation were carried out with the
Boller \& Chivens (B\&Ch) spectrograph, equipped with SITe 1k CCD.
The width of a spectrograph slit was $2.5''$ and grating was 600
lines/mm (Blue and Red), providing a dispersion of 2 \AA/pix and
with the effective instrumental spectral resolution of about 4 \AA\
($\simeq 2$\,pixels FWHM). Observations were carried out during
fotometric conditions 06--07 of February 2005, with the seeing of
$2''$ (FWHM), when the spectral range covering the Balmer lines was
observed. Exposure time for the blue part (H$\beta$ region) was
1800 s and for the red part (H$\alpha$ region) was 3000 s.

\begin{figure}
\includegraphics[angle=270,width=85mm]{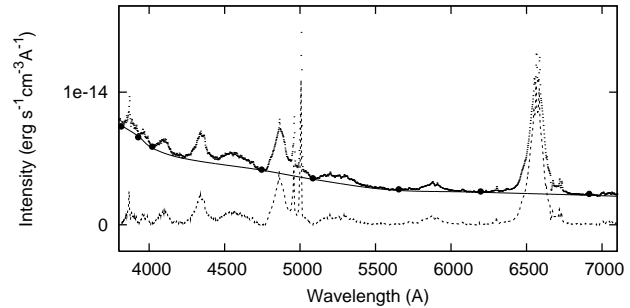}
\caption{The estimate of the continuum (solid line) in the case of
the spectrum obtained with the OAN SPM telescope. Dotted line is the
observed spectra and dashed line (below) is the spectra after
continuum subtraction. The full circles represent the points
through which the local continuum was fitted.} \label{fig01}
\end{figure}

Standard reduction procedures, including bias and flat field
corrections, wavelength calibration, spectral response, cosmic ray
hits removal and sky subtraction were performed with the
IRAF\footnotemark\footnotetext{~IRAF is the Image Reduction and
Analysis  Facility distributed by the National Optical Astronomy
Observatories, which is operated by  the Association of Universities
for Research in Astronomy (AURA) under  agreement with the National
Science Foundation (NSF). More details can be found at
http://iraf.noao.edu.} software package. The software package
DIPSO was used for reducing the level of
the local continuum (by using the DIPSO routine 'cdraw
1'\footnote{ The 'cdraw 1' procedure acts essentially as a
spline fit to the marked data points. More details for this
procedure can be found on the following website
http://www.starlink.ac.uk/.}) fitted through the  points taken
to be on the local continuum (full circles in
Figure~\ref{fig01}). When necessary, spectral lines have been
normalized to unity.

\begin{figure}
\includegraphics[width=85mm]{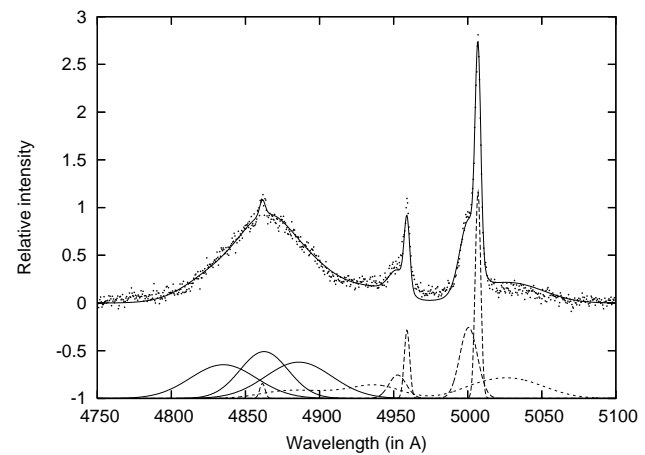}
\caption{Decomposition of the INT spectrum of H$\beta$ line of Mrk
817. The dots represent the observation and solid line is the
best-fitting. The Gaussian components are shown at the bottom. The
dashed lines at bottom represent the Fe II template, [OIII] and
H$\beta$ narrow lines. } \label{fig02}
\end{figure}

\begin{figure}
\includegraphics[width=70mm]{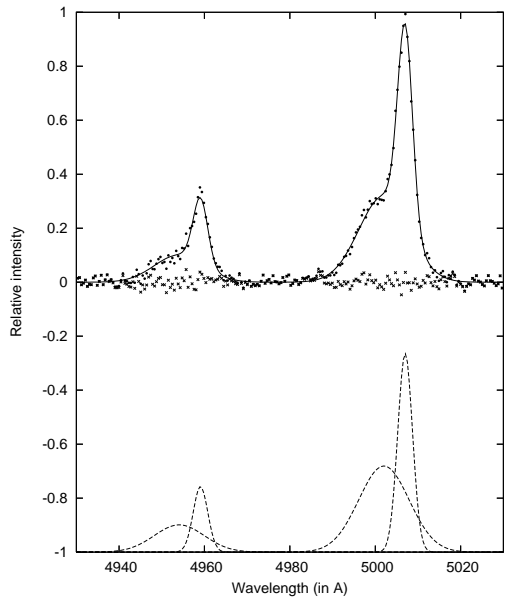}
\caption{Decomposition of the INT spectrum of [OIII]
$\lambda\lambda$ 4959, 5007 line of Mrk 817.} \label{fig03}
\end{figure}

\section{Kinematical properties of the BLR}

\subsection{Data analysis}

First, we fitted the H$\alpha$ and H$\beta$ emission lines with a
sum of Gaussian profiles\footnote{The Gaussian profile is given as
exp$[-\Delta\lambda^2/2\sigma^2]$, where $\sigma = {\rm FWHM}/2.355$
(FWHM is the Full Width at Half Maximum).} using a $\chi^2$
minimization routine to obtain the best-fitting parameters. The
fitting procedure has been described several times (Popovi\'c et
al.~2002,~2003). In the case of Mrk 817, we have assumed that the
narrow [OIII] $\lambda\lambda$4959, 5007 emission lines can be
composed by more than one Gaussian component, as the blue-asymmetry
is very obvious in their narrow lines (Popovi\'c et al.~2004). In
the fitting procedure,  we look for the minimal number of Gaussian
components needed to fit the lines, taking into account the fixed
intensity ratio of [OIII] $\lambda\lambda$ 4959, 5007 lines (the
atomic value 1:3.03) and the fit of Fe II template for the H$\beta$.
We took the relative strengths of Fe II lines from Korista (1992)
and assumed that they all originate in the same region (i.e. the
lines have the same shift and $\sigma/\lambda$). Additionally, we
assumed the core of the H$\beta$ line is originating in the same
region as Fe II lines. In order to analyse the NLR kinematics, we
fitted the [OIII] lines separately, after subtracting the continuum
and the wings of the H$\beta$ line, keeping the ratio of both
components constant ($I_{4959}:I_{5007}=1:3.03$, Figure~\ref{fig03}).
In the case of the H$\alpha$ line, for the narrow [NII]
$\lambda\lambda$ 6548, 6583 lines we assumed their intensity ratio
is 1:2.96 (Storey \& Zeippen~2000). It was found that three broad
Gaussian and one narrow components could fit well the profiles of
the H$\alpha$ and H$\beta$ lines, where we can recognize clear
evidence of substructure in these emission lines, not only in the
broad component of the lines, but also in the narrow emission lines
(Figure~\ref{fig02}).

\begin{figure}
\includegraphics[width=80mm]{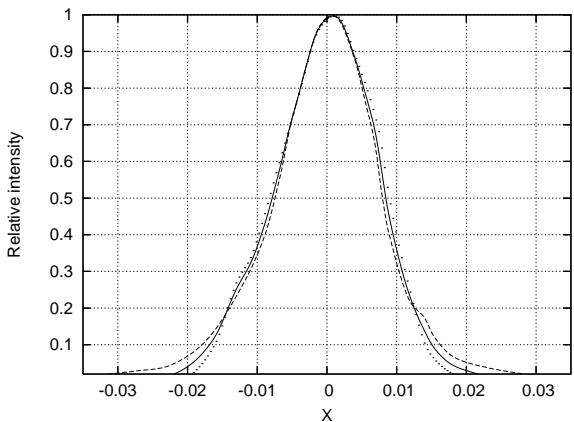}
\caption{Comparison of the normalized profiles of Mrk 817 H$\alpha$
({\it dashed line}) and H$\beta$ ({\it dots}) lines with an average
profile ({\it solid line}). The spectra have been taken with the OAN
SPM telescope. The spectral wavelengths were converted into a
velocity scale: $\lambda \to X=(\lambda-\lambda_0)/\lambda_0$. }
\label{fig04}
\end{figure}

\begin{table*}
\centering \caption{The parameters of the disc: z$_{\rm d}$ is the
shift and $\sigma_{\rm d}$ is the Gaussian broadening term
from the disc which is a measure of the velocity dispersion in the
disc, R$_{\rm inn}$ is the minimal inner radius, R$_{\rm out}$ is
the maximal outer radius. z$_{\rm G}$ and $\sigma_{\rm G}$ represent
the parameters of the Gaussian component, while F$_{\rm d}$/F$_{\rm
G}$ is the ratio of the fluxes coming from the disc and the
spherical region. p$^{min}$ is the minimal power index of the disc
emissivity, that is taken to be as
$\varepsilon=\varepsilon_0R^{-p}$, where $R$ is the distance from
the central black hole.} \label{tab01}
\begin{tabular}{@{}lccccccccc}

\hline
 $i$ & z$^{\rm min,max}_{\rm d}$  &$\sigma^{\rm min,max}_{\rm d}$ &
R$^{\rm min}_{\rm inn}$ & R$^{\rm max}_{\rm out}$ & z$^{\rm
min,max}_{\rm G}$ & $\sigma_{\rm G}$ & $p^{\rm min}$ & F$_{\rm d}$/F$_{\rm G}$\\
(\deg) & & (\kms) & $(R_g)$ & $(R_g)$ & & (\kms) & &\\
\hline
12-35 & -450,+300& 600,850 & 140 & 14000 & 0,+130 & 1100$\pm$70 &1.8 &$\sim$ 1.1\\
\hline

\end{tabular}
\end{table*}

\subsection{Results of the Gaussian analysis}

Considering only the broad components of the H$\alpha$
and H$\beta$ emission lines we can conclude that:
\begin{enumerate}
\item[(i)] the H$\alpha$ and H$\beta$ line shapes of the considered AGN are
very complex, and  usually cannot be described by one Gaussian, i.e.
the Gaussian decomposition indicates a complex kinematic structure
of the BLR.
\item[(ii)] the Gaussian decomposition indicates the existence of a
central broad component with low velocity dispersion $\sim$ 1400
\kms and a redshift consistent with the systemic redshift.
\item[(iii)] the Gaussian decomposition shows the existence of the red- and
blue-shifted broad components with higher velocity dispersion $\sim$
2200 \kms and higher (positive or negative) redshift. This implies
that the emission in the wings could originate in an accretion disc.
\end{enumerate}

As was mentioned above, we assumed that the Fe II lines originate in
the same region (with the same $\sigma/\lambda$ and shift) as the core of the
H$\beta$ line. From the best-fitting we obtained that velocity dispersion is $\sim$ 1400
\kms and that a redshift is equal to the systemic redshift of the
galaxy. With this asumption we obtained a good fit
(Figure~\ref{fig02}), therefore it could be that the Fe II emission
is created in the same region that is emitting the core of the
H$\alpha$ and H$\beta$ lines.

\subsection{Two-component model}

In order to model the BLR, we applied the two-component model where
one component is the disc or disc-like region and another one is a
spherical emission region. First component was used for fitting the
line wings and the other for fitting the line core. For the disc we
used the Keplerian relativistic model (Chen \& Halpern~1989, Chen et al.~1989).
The kinematics of the additional emission region can be described as
emission of a spherical emission region with isotropic velocity
distribution. Accordingly, the emission line profile for this region
can be described by a Gaussian function. The whole line profile can
be described by the relation:
$I(\lambda)=I_{AD}(\lambda)+I_G(\lambda),$ where $I_{AD}(\lambda)$,
$I_G(\lambda)$ are emissions of the relativistic accretion disc and
the spherical emission region, respectively (Popovi\'c et al.~2004).

For fitting the broad H$\alpha$ and H$\beta$ emission lines with the
two-component model, first we 'cleaned' the lines from its narrow
component and satellite lines, using the results from the Gaussian
decomposition (Figure~\ref{fig04}). Furthermore, we normalized the
intensities of both Balmer lines to unity and converted the
wavelengths into a velocity scale:  $\lambda \to
X=(\lambda-\lambda_0)/\lambda_0$. This allowed us to compare the
H$\alpha$ and H$\beta$ high resolution profiles. We found that they
have similar line profiles (Figure~\ref{fig04}), which
indicates that these lines are coming from the same region. For the
fitting we used the average profile of the H$\alpha$ and
H$\beta$ lines.

\begin{figure}
\includegraphics[width=115mm,angle=270]{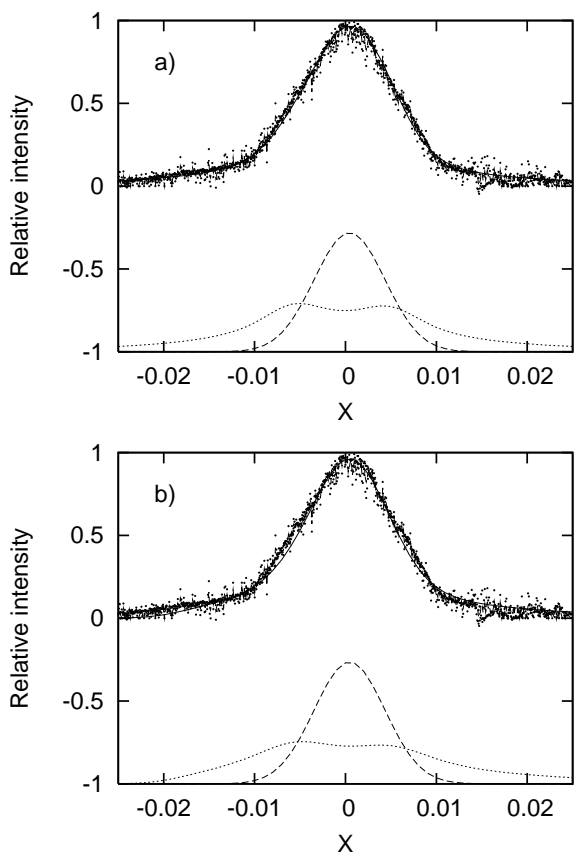}
\caption{The averaged profile of the INT H$\alpha$ and H$\beta$
emission lines fitted with the two-component model. The parameters
of the two fits are: (a) $i=30\deg$, $R_{\rm inn}=140 \ R_{\rm g}$,
$R_{\rm out}=10400\ R_{\rm g}$, $\sigma_{\rm d} = 650\ \kms$,
$p=1.8$, $\sigma_{\rm G} = 1150 \ \kms$ ; (b) $i=20\deg$, $R_{\rm
inn}=140 \ R_{\rm g}$, $R_{\rm out}=5000\ R_{\rm g}$, $\sigma_{\rm
d} = 800\ \kms$, $p=2.0$, $\sigma_{\rm G} = 1100 \ \kms$.} \label{fig05}
\end{figure}

The results of the fitting test are very dependent on the initial
values given to the parameters since we apply a two-component model
to single-peaked lines, so the number of free parameters is large.
To overcome this problem we had to use the additional constraint
that the disc component fits the line wings, and the spherical
component the line core. We have found that this model can well fit
the line profiles, but some of the parameters are not constrained.
The unique solution for the model could not be obtained without
giving additional constrains to the disc parameters, but we were
trying to find the optimal solution. This is presented in the
Figure~\ref{fig05} were two satisfactory fits, obtained with
different sets of parameters, are shown. From these fitting tests
(without any constraints for the disc parameters) we were able only
to give rough estimates of the kinematical parameters of the BLR in
the AGN Mrk 817 (see Table~\ref{tab01}). Important result is that
the ratio of the disc flux and the flux from the spherical region is
close to unity and approximately {\bf the same} in every fit. Obviously,
in this AGN, the shape of the line wings indicates rotational
motion, which may be caused by a disc-like geometry.

\begin{figure}
\includegraphics[width=78mm]{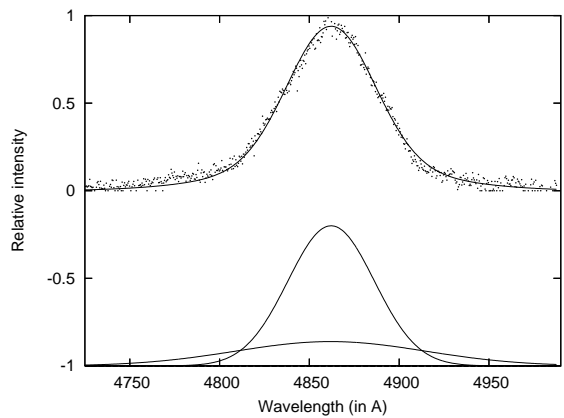}
\caption{ The broad profile of the INT H$\beta$ emission line
fitted with two Gaussian functions.} \label{fig06}
\end{figure}

We should note here that other interpretations for the structure of
the complex broad emission line region are possible. For instance a
Very Broad Line Region (VBLR), that has isotropic velocity
distribution, may be responsible for the emission in the line wings.
Therefore, we performed an additional test, in what we fitted the
average profile of the broad H$\alpha$ and H$\beta$ emission lines
with two Gaussians. We obtained a relatively good fit (see
Figure~\ref{fig06}) in the case when one Gaussian fits the line core
(with velocity dispersion $\sim$ 2100 \kms), and the other one fits
the line wings (with velocity dispersion $\sim$ 4600 \kms). We
should note here that the BLR might be composed from two spherical
regions, but we should emphasize that we obtained better fit with
the disc + spherical emission region model (see Figure~\ref{fig05}).
This indicates that probably in the case of Mrk 817 the part of the
BLR emission is coming from the disc-like region.

\section{The Narrow Line Region}

The Narrow Line Region (NLR) of the studied AGN also shows a complex
structure, and we can clearly see at least two NLR regions (see Figure~\ref{fig03}):
\begin{enumerate}
\item[(i)] the NLR1, which has $\sigma \sim$ 450 \kms,
and relative approaching velocity 400 \kms with respect to
the systemic redshift of the observed galaxy; and

\item[(ii)] the NLR2 which has $\sigma \sim$ 150 \kms,
and a redshift equal to the systemic one of the studied
object.
\end{enumerate}

The blue asymmetry observed in the [OIII] lines of Mrk 817 is
also detected in numerous AGN, especially the case that the
blue-shifted component is broader (see e.g. Veilleux 1991, Leighly
1999, Leipski \& Bennet 2006). The difference in shifts and widths
of the [OIII] $\lambda\lambda$4959, 5007 between two narrow
components indicate different kinematical and physical properties of
these two NLRs. The clear tendency of the NLR1 to have a
blue-shifted systemic velocity suggests that it is associated with a
high-ionization outflow originating in the accreting source. In this
case the receding jet emission in the [OIII] lines might be obscured
or absorbed by the host galaxy, so one sees only the outflowing gas
from  the closer part of the jet.

\begin{table*}
\centering \caption{Flux ratio of the Balmer lines of the AGN Mrk
817: (a) measured values; (b) values corrected for the intrinsic
reddening. The flux of the H$\beta$ line is given in ${\rm erg \
cm^{-2} \ s^{-1}}$.} \label{tab02}
\begin{tabular}{@{}lcccccc}
\hline
 & F$_{\rm H\alpha}$/F$_{\rm H\beta}$ & F$_{\rm H\gamma}$/F$_{\rm H\beta}$ &
  F$_{\rm H\delta}$/F$_{\rm H\beta}$
 & F$_{\rm H\epsilon}$/F$_{\rm H\beta}$ & F$_{\rm H\beta}$ \\

\hline
(a) & 3.289$\pm$ 0.750 & 0.371$\pm$0.092 & 0.183 $\pm$0.046  & 0.088$\pm$0.027 & (3.120$\pm$0.361)E-13\\

(b) & 2.579$\pm$ 0.588 & 0.405$\pm$0.100 & 0.208 $\pm$0.052 & 0.101$\pm$0.031  & (3.120$\pm$0.361)E-13\\

\hline

\end{tabular}
\end{table*}

\section{Balmer line intensities as indicators of the BLR physics}

Here we have used the  strong Hydrogen lines to discuss the physical
properties of the Mrk 817 BLR. Assuming that the BLR is optically
thin (see Corbin \& Boroson~1996)\footnote{Corbin \& Boroson (1996)
found from the difference between the Ly$\alpha$ and H$\beta$ full
width at zero intensity (FWZI) values additional evidence of an
optically thin BLR.} we have applied the method (so called Boltzmann
Plot - BP) proposed by Popovi\'c (2003) on the flux ratios of the
Hydrogen lines of the Balmer series (H$\alpha$, H$\beta$, H$\gamma$,
H$\delta$, and H$\epsilon$), obtained with the OAN SPM telescope.
Before measuring the emission line flux, it was necessary to 'clean'
the lines from its narrow component, as well as from the satellite
lines (especially in the case of the H$\alpha$ and H$\beta$). In the
case of the H$\gamma$, the Fe II contribution was subtracted using
the Fe II template of V\'eron-Cetty et al.~(2004). The flux ratios
of the Balmer lines are given in the Table~\ref{tab02}. The errors
were estimated as cumulative errors of the continuum subtraction
($\sim 10\%$) and the line flux measurement. The errors of the
narrow and satellite lines subtraction are also present, but we
estimated that errors of this procedure are within the frame of the
above errors.

Also, we took into account the reddening effects. In the case of Mrk
817, the influence of the galactic reddening is insignificant,
according to data in the NASA's Extragalactic Database
(NED)\footnote{http://nedwww.ipac.caltech.edu/}, but the intrinsic
reddening is higher, E(B-V)=0.22 (Cohen 1983) and should be taken
into account. The values of the unreddened flux are also given in
the Table~\ref{tab02} (as value (b)). This reddening
coefficient has been determined for the NLR which implies that the
reddening could be even higher in the BLR, and consequently the
obtained electron temperature may represent the lower limit.

\begin{figure}
\includegraphics[width=48mm,angle=270]{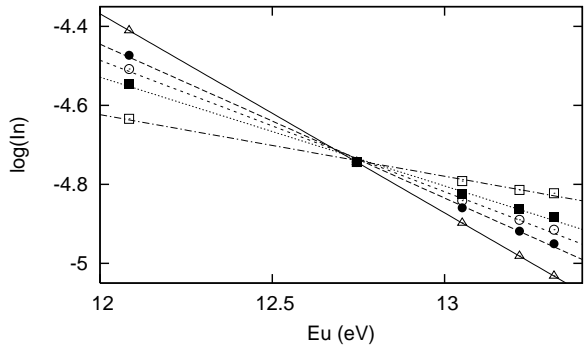}
\caption{ The BP for the BLR composed of the one region in the
PLTE and one that emits recombination lines ("Case B"
recombination). Notation: open triangles represent pure PLTE for $T=10000$ K, open
squares represent pure recombination ($T=32300$ K from the BP), full circles for
$F_{PLTE}/F_{non-PLTE} = 2$ ($T=12900$ K from the BP), open circles
for $F_{PLTE}/F_{non-PLTE}=1 $ ($T=15200$ K from the BP), full squares
for $F_{PLTE}/F_{non-PLTE} = 0.5$ ($T=18300$ K from the BP) .}
\label{fig07}
\end{figure}

Also, it seems that the BLR of Mrk 817 is composed from two
kinematically, and probably physically different regions (see \S
3.3), and the BP is applied on the summary emission of these two
regions. In Popovi\'c et al. (2006) several combination were
demonstrated, where the BLR is composed of the one region in the
Partial Local Thermodynamical Equilibrium (PLTE) and the other one
in the non-PLTE conditions, and here we present one example. In this
example, we assumed that in one region PLTE is present with
$T=10000$ K, and that other one, non-PLTE region, emits
recombination Balmer line spectrum ("Case B") with $T_e=10000$ K,
$N_e=10^6\rm cm^{-3}$ (Osterbrock 1989, Table 4.4). We considered
that the contribution of each region in the total emission flux
($F_{tot}$) can be different taking that
$$F_{tot}={F_{PLTE}+p\times F_{non-PLTE}\over{1+p}},$$
where $F_{PLTE}$ and $F_{non-PLTE}$ are the flux of the region in
PLTE and non-PLTE, respectively. The coefficient $p$ is varied from
0.5 to 2 so different contributions of the non-PLTE
emission in the total line flux were calculated (the $F_{H\beta}$ is
taken to be 1). In the Figure~\ref{fig07} the BP is presented for
the different values of the parameter $p$, as well as for the cases
when only PLTE or non-PLTE is present. As one can see from
Figure~\ref{fig07}, the determined temperature for a
mixed BLR (one part in PLTE and one in the 'Case B' recombination) are higher
than they really are in the assumed different parts of the BLR. As it was
shown in Popovi\'c et al. (2006), if the BLR temperature obtained from BPs are
smaller than 20000 K, one may expect that PLTE exist at least in one part
of the BLR.

\begin{figure}
\includegraphics[width=46mm,angle=270]{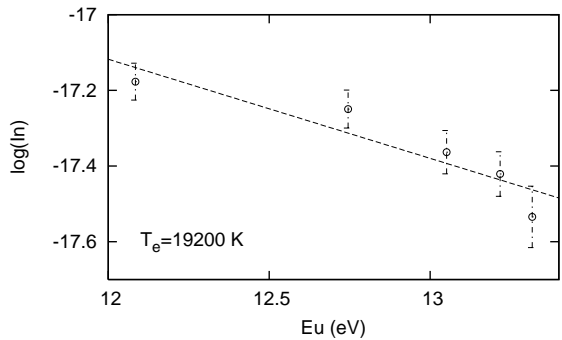}
\caption{The BP applied on the Balmer lines of the AGN Mrk 817,
obtained with the OAN SPM telescope, after subtracting the influence
of the intrinsic reddening. The corresponding temperature, $T_{\rm
e}=\log(e)/kA$, is given in the upper left corner of the plot.}
\label{fig08}
\end{figure}

For Mrk 817, in both cases, with or without reddening taken into
account, the BP could be applied, meaning that the flux ratio of the
Balmer lines could be fitted with the straight line, but with
different slopes. On the Figure~\ref{fig08}, log(In) vs.
E(u)\footnote{The log(In) is logarithm of the spectrally integrated
emission line intensity and is given as $\log(I_n)=\log{I_{ul}\cdot
\lambda\over{g_u A_{ul}}}$, where $I_{lu}$ is relative intensity of
transition from upper to lower level ($u\to l$), $g_u$ is the
statistical wight of the upper level and $A_{ul}$ is the transition
probability. The E(u) is the energy of the upper level of the line
transition (Popovi\'c~2003).} is presented for the case where
correction for the intrinsic reddening was performed. According to
calculated slopes, one may consider that PLTE exists at least in one part of
the BLR, and consequently give estimates of the electron temperature
(Popovi\'c~2003). Using the relationship given by Popovi\'c (2003)
and a temperature parameter $A$ obtained from the slope (see
Figure~\ref{fig06}) we found that the electron temperature
might be in an interval of $10000 < T_{\rm e} < 20000 $ K, depending
on the reddening effect and the contribution of different regions to
the total flux of broad Balmer lines.

\section{Structure of the emission line region of Mrk 817}

\begin{figure}
\includegraphics[width=84mm]{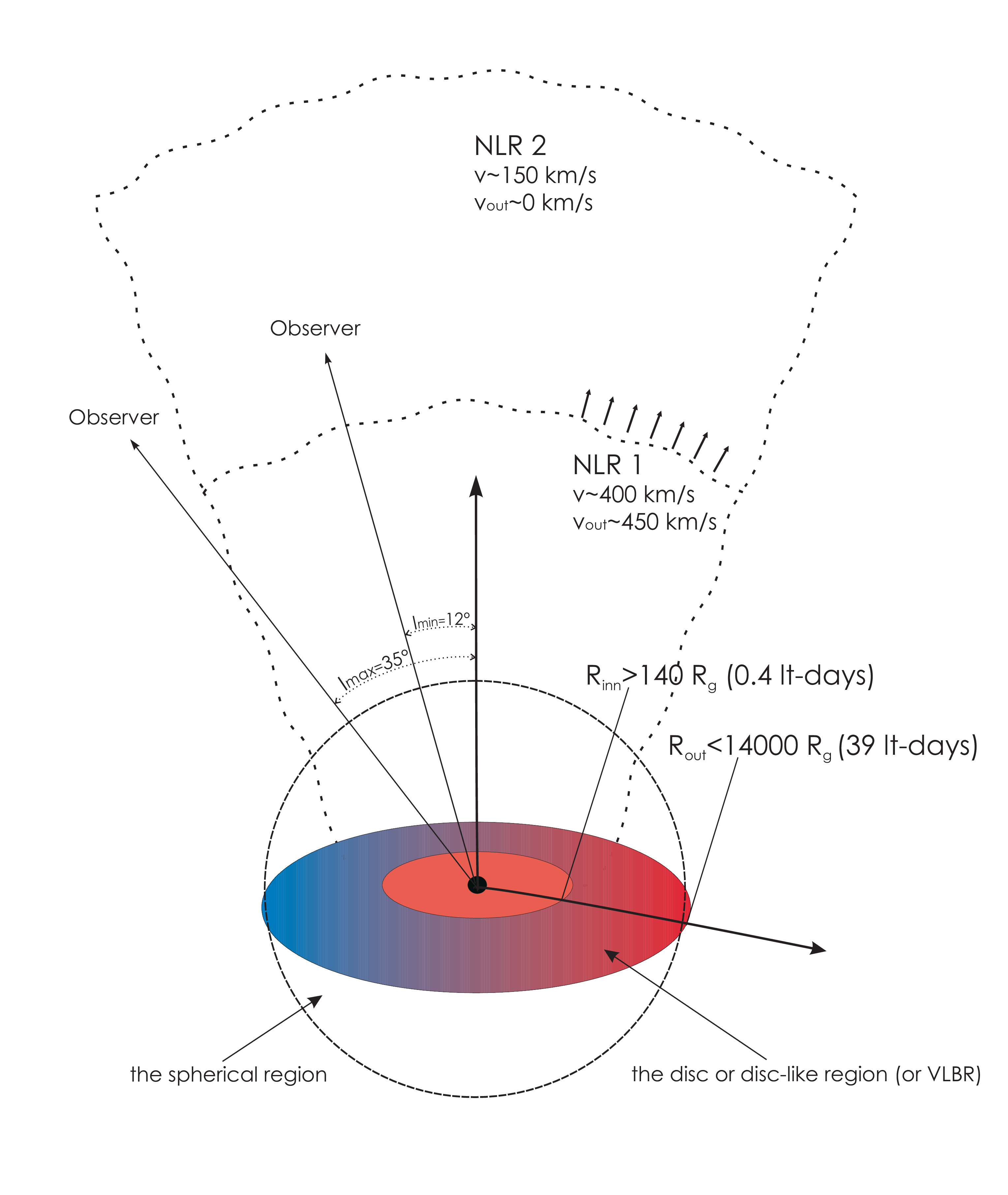}
\caption{The scheme of the {\bf complex} BLR and NLR of the AGN Mrk
817. The radius in light days was calculated for the black hole mass
of $4.9 \times 10^7 M_{\odot}$. $v_{out}$ is the outflow velocity
and $v$ is the  velocity dispersion obtained from the FWHM.}
\label{fig09}
\end{figure}

>From the study of the narrow and broad emission lines we can make
some conclusions on the emission line region structure of Mrk 817.
In general two main emission regions (NLR and BLR) are present, but
they seem to be complex as well. A scheme of the emission line
region of Mrk 817 is presented in Figure~\ref{fig09} and it highly
corresponds to the unified model scheme.

In the inner part of the emission region probably a relativistic
disc or a disc-like region is present (or VLBR see \S3.3). Our
analysis of the disc parameters shows that the minimal inner radius
of the disc cannot be smaller than 0.4 lt-days and that the disc
should be smaller then 39 lt-days (for a black hole mass of $4.9
\times 10^7 M_{\odot}$). These results are in the agreement with the
results obtained by Kaspi et al. ~(2000), who estimated the
dimensions of the BLR to be $\sim 15$ lt-days. Also, the inclination
of the disc is small and in the range of $12 \deg < i < 35 \deg$.
The other region seems to be spherical with isotropic velocity
distribution. It is hard to estimate its dimensions, but one can
conclude that it should not be significantly larger than the disc,
since the  velocity dispersion in this region ($\sim 10^3 \ \kms$)
is very similar to the velocity dispersion in the
disc\footnote{Beside the rotational motion, one can expect the
random motion inside of the disc, therefore we adopted the model for
the disc, proposed by Chen \& Halpern (1989), where random motion is
also taken into account.} (see Table~\ref{tab01}).  If we assume
that the random velocities are decreasing from the center of an AGN,
one may conclude that these two regions are not too different in
dimensions. It is therefore possible that this region is created by
the disc wind (Murray \& Chiang~1997).

In this complex BLR, it is hard to determine the temperature
(see \S5), but from the BP one may have an impression that the
temperature might be in an interval from 10000 K to 20000 K. As it
was mentioned above, we could not use the BP method as an indicator
of the physical parameters of the disc (or VBLR) and the spherical
region, since we did not have high resolution observation of the
H$\gamma$, H$\delta$ and H$\epsilon$ lines, but the BP indicates
that at least one of the two regions might be close to the PLTE (see
Popovi\'c et al. 2006).

The narrow line region is also complex and it consists of the NLR1,
with $\sigma\sim$ 450 \kms, and relative approaching velocity 400
\kms with respect to the systemic redshift of the observed galaxy,
and the NLR2, with $\sigma\sim$ 150 \kms, and a redshift equal to
the systemic one of the corresponding object. Since the inclination
angle of the disc is relatively small it is likely that we observe
this AGN along the approaching jet that emits only in the NLR1.
Therefore, the outflow velocities should be higher for the factor of
$\sin i$. There is no clear connection between the spherical region
and NLR1, but one can conclude that emission of the approaching jet
in the optical spectral band is noticeable only far from the BLR. It
is, therefore, interesting that we cannot register an outflow in the
spherical part of the BLR. It does not mean that we have no outflow
in this region, but it might be that in this part the emission of the
jet is in the radio spectral band. The high-resolution VLA map of
Mrk 817 at 8.4 GHz (Kukula et al. 1995) shows a relatively symmetric
shape of the radio emission, that may come from a line-of-sight
orientated jet.

\section{Conclusions}

In this contribution, we have reported the results of the emission
line regions study of the Seyfert 1.5 galaxy Mrk 817. The Seyfert
1.5 galaxies are convenient for exploring the narrow and broad line
emission regions since those galaxies have very  strong both narrow
and broad emission lines. In our work we have used several sets of
different spectral observations with high spectral resolution. Our
main finding is that both NLR and BLR are complex emission regions,
where they are composed of at least two kinematically separated
regions. We made a scheme of the emission line regions of Mrk 817
(Figure~\ref{fig09}).

We found that the BLR of Mrk 817 can be described with the
two-component model, where the core of the line is coming from a
spherical region with isotropic velocity distribution, and wings are
probably affected by a low inclined accretion disc. {\bf The ratio
of the fluxes coming from the disc and from the spherical region did
not depend on the parameters of the fit, it was almost the same in
evert fit and equal to unity}. In the NLR, we found that an outflow
is present. In principle we can conclude that emission line region
of Mrk 817 is complex, heaving at least four kinematically and
physically different regions.

\section*{Acknowledgments} This work was supported by the Ministry of Science
and Environment Protection of Serbia through the project
``Astrophysical Spectroscopy of Extragalactic Objects''. L. \v C. P.
is supported by Alexander von Humboldt Foundation through
"R\"{u}ckkehrstipendium". V. H. C. is supported by the CONACYT
research grant 39560-F (Mexico). We are grateful to Nina Polusuhkina,
who provide us spectra of Mrk 817 observed by K.K. Chuvaev and to
Professor Philippe V\'eron for useful comments and suggestions.

\label{lastpage}
\end{document}